# Adaptive Point-to-Multipoint Transmission for Multimedia Broadcast Multicast Services in LTE

Mai-Anh Phan, Jörg Huschke
Ericsson GmbH
Herzogenrath, Germany
{mai-anh.phan, joerg.huschke}@ericsson.com

**This paper investigates point-to-multipoint (PTM) transmission supporting adaptive modulation and coding (AMC) as well as retransmissions based on incremental redundancy. In contrast to the classical PTM transmission which was introduced by the Multimedia Broadcast Multicast Service (MBMS), the adaptiveness requires user individual feedback channels that allow the receivers to report their radio conditions and send positive or negative acknowledgments (ACK/NACK) for a Layer 1 transport block to the eNodeB. In this work, an adaptive PTM scheme based on feedback from multiple users is presented and evaluated. Furthermore, a simple NACK-oriented feedback mechanism is introduced to relieve the feedback channel that is used in the uplink. Finally, the performance of different single-cell MBMS transmission modes is evaluated by dynamic radio network simulations. It is shown that adaptive PTM transmission outperforms the conventional MBMS configurations in terms of radio resource consumption and user satisfaction rate.**

## I. Introduction

In the past few years an increased data traffic usage has been observed in mobile networks. One of the key data services is mobile TV whose popularity is significantly rising worldwide. Due to the growing importance of data services the Third Generation Partnership Project (3GPP) launched a work item for Global System for Mobile Communication (GSM) and Universal Mobile Telecommunication System (UMTS) known as the Multimedia Broadcast Multicast Service (MBMS). The MBMS specifications [1], [2] in UMTS Release 6 were finalized in 2005. MBMS was developed in order to support a more efficient delivery of identical multimedia contents to several consumers interested in the same data service. With the introduction of a new point-to-multipoint (PTM) bearer, the unicast solution in UMTS was extended by broadcast capabilities, thus enabling an unlimited number of users to simultaneously receive the same data service on common radio resources. While a feedback channel is provided for each point-to-point (PTP) connection allowing for fast user individual link adaptation, PTM transmissions are unidirectional and therefore statically configured with a fixed modulation and coding scheme (MCS) as well as constant power such that a certain coverage probability is reached. If neighboring cells offer the same MBMS service, the link performance can be enhanced by macro diversity which allows for soft combining of several radio links. However, this strategy is only efficient for mobile mass-media services.

If the user density is so low that there are cells without MBMS users, PTM transmission from a cluster of cells would obviously lead to a waste of radio resources. If only a small user group is addressed, it is most likely that PTM transmission is too robust, thus utilizing more radio resources than necessary. MBMS therefore allows for data delivery over PTP connections for a very low number of receivers. In order to determine the number of interested users in a cell, 3GPP has also specified a so-called counting procedure for MBMS. Furthermore, repeated counting also supports dynamic switching between PTP and PTM bearers depending on the number of users requesting the same MBMS service in a cell. Thus, the network can adapt to the actual user requirements providing MBMS with increased flexibility. Nevertheless, the optimal break-point between both MBMS delivery modes strongly depends on the users' instantaneous radio conditions.

Since the probability that robust transmission is required decreases with smaller group size, the MBMS performance can be further enhanced if link adaptation is applied. With this MBMS feature, the transmission can be adjusted to fulfill the actual radio requirements of the user group. In contrast to classical broadcast technologies lacking an uplink it is easy to integrate a feedback channel in cellular networks in order to support adaptive modulation and coding (AMC) and enable retransmissions.

In case of download services in MBMS, application layer coding can be adopted. However, file repair techniques require rather large data volumes to be efficient resulting in larger delays. As the scope of this study is the enhancement of reliable data transmission for time-critical services such as mobile TV, which only tolerate moderate delay, application layer coding is not investigated.

In this paper, an adaptive PTM transmission method, which is based on receiver feedback including channel quality reports and positive/negative acknowledgments of received packets, is presented. For link adaptation purposes different techniques on the processing of channel quality and packet reception reports from multiple receivers are introduced. Afterwards, the system model is described that is used for the radio network simulations. Subsequently, the performance of adaptive PTM transmission for mobile TV services is illustrated and compared with conventional PTP and PTM transmission methods. Finally, a conclusion is drawn based on the simulation results.

## II. POINT-TO-POINT TRANSMISSION IN LTE

The use of Orthogonal Frequency Division Multiplex (OFDM) [3] for downlink transmission in Long Term Evolution (LTE) of UMTS enables flexible frequency domain scheduling. In each transmit time interval (TTI) of 1 ms, one or more so-called subbands with a granularity of 180 kHz corresponding to 12 subcarriers can be assigned to a user equipment (UE). The number of available subbands depends on the total downlink system bandwidth. Based on reference symbols the UE can measure the reception conditions in each subband. In basic configurations of PTP transmission, active UEs periodically report their reception conditions as channel quality indicators (CQI) to the serving eNodeB. This concept is referred to as multiband CQI reporting because it provides channel quality information for multiple subbands, thus allowing for channel-dependent scheduling. This means that subbands indicating good channel qualities are assigned to the corresponding UEs first, aiming at the maximization of the transmit rate. Consequently, the bandwidth and frequency range assigned to a UE may vary. The appropriate MCS for the group of subbands is dynamically selected based on the CQIs aggregated in the frequency domain.

Due to inaccurate measurements, user mobility and traffic fluctuations, rate adaptation errors may occur resulting in either unnecessarily high robustness of the transmission or Layer 1 block errors. In order to compensate for inaccuracies of AMC, an N-channel stop-and-wait Hybrid Automatic Repeat reQuest (HARQ) protocol [4] has been introduced in LTE. If the decoding of a transport block was successful, the receiver sends a positive acknowledgment (ACK) to the transmitter; otherwise it sends a negative acknowledgment (NACK). Information from previously failed attempts is saved in a buffer and soft combined with successive information before a decoding attempt is started. Consequently, the transport block becomes more reliable with each retransmission. Especially in environments with rapidly changing channel conditions, HARQ significantly improves the system performance.

Since there is a fixed time alignment between the transmission of a transport block and the reception of a HARQ feedback, only a single bit has to be allocated for a HARQ status report.

## III. ADAPTIVE POINT-TO-MULTIPOINT TRANSMISSION

This paper aims at combining the advantages of broadcast and unicast transmission. A novel approach for the enhancement of broadcast/multicast strategies is to exploit feedback information for PTM transmission similar as used for dedicated channels. Feedback about channel quality and HARQ status allow for link adaptation of PTM transmission, thus improving efficient resource allocation. While the transmitted signal can be adapted to the aggregate instantaneous channel quality, the data must only be transmitted once, addressing multiple MBMS users simultaneously. Outer ARQ in the Radio Link Control (RLC) layer to increase the reliability is not considered because it causes additional delay and overhead. Furthermore, the small residual error rate after HARQ error correction can be tolerated by mobile TV applications. Thus, RLC is applied in unacknowledged mode (UM).

In the baseline scenario of this work, dedicated feedback channels in the uplink are allocated to each MBMS user to carry CQI and HARQ status reports. The feedback channels are identical to those used in PTP transmission in order to maximize the reuse of existing protocols and functionalities used for the PTP scheme. In contrast to user individual MCS selection, the new scheme requires preprocessing of multiple CQI reports before choosing an appropriate MCS for a single downlink transmission. Concerning HARQ status reports, multiple uplink resources have to be scheduled in response to the transmission of a single downlink transport block. The advantage of this scheme is that it is identical to PTP as soon as there is only one single user in the MBMS group. With an increasing number of users in the MBMS group, though, the uplink traffic caused by feedback would rise accordingly. However, uplink load may be reduced by omitting redundant information.

### *A. Adaptive Modulation and Coding in MBMS*

In PTM with feedback, the eNodeB may receive channel quality information from multiple users in the MBMS group, but the MCS is chosen only for a single downlink transmission resource which is commonly received by all users interested in the concerned MBMS service. Consequently, preprocessing of all received CQIs is required to obtain an aggregate channel quality for AMC. In general, it is desired to offer Quality of Service (QoS) to the majority of the users. Therefore, it is reasonable to adapt PTM transmission to the user which experiences the lowest channel quality. This implies that the chosen MCS is excessively robust from the point-of-view of other receivers in the MBMS group. Furthermore, the robustness of the transmission also covers the possibility that the channel quality of another user may deteriorate due to multipath fading.

In Fig. 1, the probability density functions (PDF) of the channel quality experienced by the worst user are depicted for different group sizes. The channel quality is measured by the normalized Signal-to-Interference-plus-Noise-Ratio (SINR), assuming a transmit power of 1 W. The solid line shows the channel quality of single users as a reference. The graph shows that the probability of having a user with poor channel quality increases with the MBMS group size.

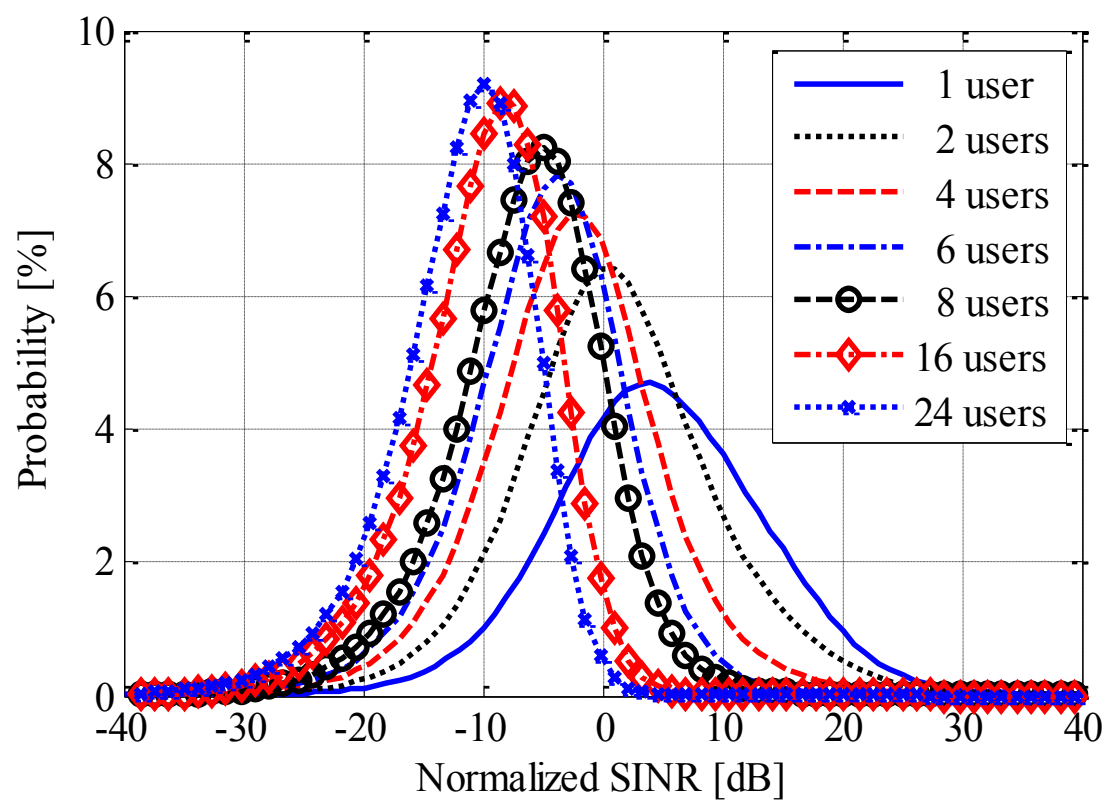

Figure 1. MBMS worst case channel qualities for different group sizes and a system load of 70 %.

In order to avoid that a single user with temporarily poor channel quality significantly deteriorates the overall throughput of the group, the following minimum-CQI time-domain (TD min-CQI) scheduling technique is introduced. A threshold for a minimum channel quality to achieve a minimum throughput per transmission is set. As soon as there is a user with channel conditions below the threshold, the whole MBMS group is not scheduled in the corresponding TTI. Otherwise, the MCS is selected based on the worst user.

### *B. Hybrid Automatic Repeat Request (HARQ) in MBMS*

Similar to PTP transmission, the HARQ technique can also be applied for MBMS. In [5], different error control algorithms for multicast delivery have been investigated for UMTS. Compared to selective repeat ARQ, a HARQ scheme based on transmission groups (TG) of multiple packets reduces the feedback from per-packet to per TG-feedback and leads to higher throughput as the number of users per multicast group increases. In [6], the HARQ scheme for multicast delivery is investigated for HSDPA and is based on single packets rather than TGs. The HS-DSCH is used for PTM transmission with fixed MCS and power, but the number of retransmissions is varied based on HARQ feedback. Compared to a fixed number transmissions per transport block, HARQ feedback permits resource savings of 70 % for 2.38 users per cell and 40 % for a group size of 23.8 users.

According to the 3GPP specifications for PTP HARQ scheme in LTE, a combined ACK and NACK feedback syntax is used to indicate to the transmitter successful or failed decoding of packets, respectively. In PTM transmission mode, a HARQ retransmission is triggered if the transmitter receives at least one NACK from the receivers as long as the maximum number of transmission attempts has not been reached. Obviously, the uplink traffic caused by HARQ status reports increases linearly with the number of users. However, it is sufficient to operate on an exclusive NACK syntax to reduce the uplink overhead. For this scheme, it is sufficient for the transmitter to detect that at least one receiver has failed to decode the transport block correctly to trigger a retransmission. It should be noted that a combined ACK and NACK feedback scheme is beneficial if a UE does not receive the scheduling information. In such a case, the UE is not aware of the transmission and thus, it does not send any feedback information. In a combined ACK and NACK scheme, such a failure can be detected by the transmitter. In an exclusive NACK scheme, the absence of a feedback would be misinterpreted as a successful transmission, although the UE has not received the data. However, the probability of missing scheduling information is rather low and can therefore be tolerated by mobile TV.

Since HARQ feedbacks only need to allocate a single bit, a common feedback channel can be used in case of an exclusive NACK scheme. Thus, no additional resources for HARQ status reports are needed compared to the case of dedicated PTP transmission. If multiple receivers access the common channel at the same time to transmit a NACK signal, the signals can be superimposed either constructively or destructively. In case of a collision, the feedback channel must provide the property that the eNodeB is still able to detect that at least one NACK has been sent. The eNodeB measures the receive power in the resource dedicated for HARQ feedback transmission. If the receive power exceeds a certain threshold it is assumed that at least one NACK was transmitted. The threshold value allows trade-offs between reliability and throughput. Similar to avoid misinterpreting NACKs as ACKs or vice versa, it is required that the probability for the detection of NACKs is sufficiently high, especially because RLC UM is applied. As a consequence, the threshold must be chosen sufficiently low.

### *C. NACK-Oriented Feedback Scheme in MBMS*

It is desirable to keep uplink transmissions at a minimum level to save battery power in the UE. Additionally, reduction in uplink transmission implies that eNodeBs receiving data from users with high uplink traffic data rates experience a decreased interference level. Periodic CQI reporting can become highly inefficient because multiband CQI reporting, which gives information about the channel quality of each subband, consumes a relatively high amount of resources. As the MCS is selected based on the lowest CQI values, other CQI reports are redundant and therefore simply discarded. Consequently, redundant feedback information needs to be identified in order to avoid transmitting unnecessary feedback. In general, if the MCS is not sufficiently robust for a user in the MBMS group, it will fail to decode the transport block. Block errors are simple means to indicate that the MCS needs to be adjusted. Therefore, a so-called NACK-triggered CQI reporting may be introduced. This means that CQI values are only transmitted when a NACK signal is sent. Consequently, this approach reduces combined ACK/NACK to exclusive NACK feedbacks and periodic to NACK-triggered CQI reporting. The adaptation of the MCS can be divided into different phases: the actual adaptation phase and the recovery phase. In the actual adaptation phase the MCS is chosen according to the aggregate CQIs that have been reported together with the NACK signal. In the recovery phase the robustness of the MCS is gradually reduced with every transmission containing new data resulting in a blind or passive adaptation. The intensity of the recovery steps may be set according to the success rate.

## IV. System Model

A high complexity radio network simulator with propagation and interference models is used to evaluate the gains achieved by various adaptive PTM transmission techniques compared to conventional PTP and classical PTM transmission using a fixed MCS. In this work, the mobile TV service with a source data rate of 128 kbps is delivered to MBMS users by applying the previously described transmission schemes. All video frames have equal size to simplify both the quality model and the configuration for PTM transmission with fixed MCS. Error-free channel quality measurements and CQI reporting are assumed in the simulations. In contrast, ACK-to-NACK and NACK-to-ACK errors occur with a probability of $10^{-3}$. SINR values are averaged over one TTI. Instantaneous SINR is mapped to mutual information [8] in order to obtain the block error probability. The MCS is selected such that the mutual information is maximized, aiming at a target BLER of 10 % per user in both PTP and adaptive PTM transmission mode.

## V. SIMULATION SCENARIO

The simulated radio network consists of seven sites, which cover three sectors each, thus resulting in 21 hexagonal cells. A wrap-around scheme is used to avoid border effects. Further parameters concerning the radio network model and the simulation parameters are summarized in Table I.

TABLE I. SIMULATION PARAMETERS

| | |
|---|---|
| Cell layout | 3-sector hexagonal grid |
| Network size | 7 sites → 21 cells |
| Cell radius | 500 m |
| Distance attenuation | 29.03 + 3.52 · 10 · log(d)<br>distance d in [m] |
| Multipath fading | 3GPP Typical Urban |
| Shadow fading | log-normal, σ = 8 dB |
| Spectrum allocation | 5 MHz |
| eNodeB transmit power | 20 W / 13 dBW |
| Maximum antenna gain | 14 dBi |
| Receiver | Single antenna,<br>no receiver diversity |
| UE noise factor | 8 dB |
| UE speed | 3 km/h |
| UE CQI reporting period | 10 ms |
| Max. no. of HARQ transmissions | 8 |
| TD min-CQI threshold | $5^{th}$ percentile of channel quality |

Each simulation contains a single MBMS session which the UEs with different life times may join and leave at arbitrary time instances. However, the number of UEs that are concurrently in the simulation area is kept constant, so that the average number of MBMS users per cell is also fixed. When a UE terminates its session and leaves the system, a new UE is created. In total, each simulation run creates 10 000 UEs. Since UEs are generated at random positions throughout the whole simulation area according to a uniform distribution, the actual number of users per cell may vary. Furthermore, the UEs move with a speed of 3 km/h.

In order to achieve fair comparison, the conditions must be the same for all scenarios. Therefore, the traffic is chosen such that the system load reaches approximately 70 %. There are PTP users in each cell which are responsible of creating background traffic. The performance of the background users is not considered in this work, which means that the impact on background users caused by MBMS users is not investigated.

### A. Scheduling

Scheduling decisions are based on different weights. In PTP transmission mode, channel-dependent scheduling is applied, i.e. UEs with high channel quality are prioritized. In order to avoid that MBMS users with poor radio conditions are not served in PTP transmission mode, a packet age weight is used for MBMS users in the unicast scenario. In order to prevent that extremely late packets are transmitted to the client and thus, waste valuable resources, old packets are dropped.

### B. Performance Metrics

The comparison between different kinds of transmission modes is based on the average resource allocation per user and the user satisfaction rate (USR), which is also referred to as coverage. In LTE, power control plays a minor role. Instead, subbands are assigned to UEs in each TTI. It can be assumed that the total transmit power is evenly distributed to all available subbands. Consequently, power consumption as a measure for resource allocation can be used. As HARQ retransmissions are supposed to improve system performance, the average number of HARQ transmission attempts, which are necessary for the receivers to correctly decode a transport block, is also analyzed.

### C. User Satisfaction

There are various parameters that impact the perceived quality of streaming video. These parameters include the data rate of the video, the initial buffering time, and the fraction of lost video frames. Video frames may get lost, which can be caused by poor radio conditions, dropping of old packets, or simply late arrival. Since the number of HARQ transmission attempts is restricted, there may be receivers that have not been able to decode the packet after the maximum number of transmission attempts. In case of late arrival, the video frame arrives after it is supposed to be played out, and thus, it is also regarded as being lost. In this model, a user can either be satisfied or unsatisfied. Here, it is assumed that the user is willing to wait for 0.5 seconds until video start before becoming unsatisfied. The user expectations cannot be fulfilled, either, when the accumulated buffering time exceeds 0.5 seconds. Furthermore, a video loss rate of 1 % is tolerated. It is required that a user satisfaction rate of at least 95 % is reached for all modes. Note that the PTP and adaptive PTM schemes aim at a higher coverage, though.

## VI. SIMULATION RESULTS

The delivery of the mobile TV service to multiple receivers applying different single-cell MBMS transmission modes is investigated by radio network simulations. The following section presents and evaluates the obtained simulation results. As only one MBMS service is simulated, the number of users per cell corresponds to the group size. Fig. 2 depicts the average power consumption per MBMS group for different group sizes. If PTP transmission mode is used, the power consumption increases almost linearly with the group size. For the PTP scheme, minor gains can be achieved with larger group sizes because channel-dependent scheduling can be more efficiently exploited. As the power settings and MCS selection for classical PTM without feedback are fixed, the resource allocation per cell is constant, while the power consumption per user decreases proportionally with the group size. The power gain is defined as

$$P_{gain} = \frac{P_{ref} - P_0}{P_{ref}} \quad (1)$$

where $P_0$ is the considered average power consumption and $P_{ref}$ is the power consumption needed in the reference scenario.

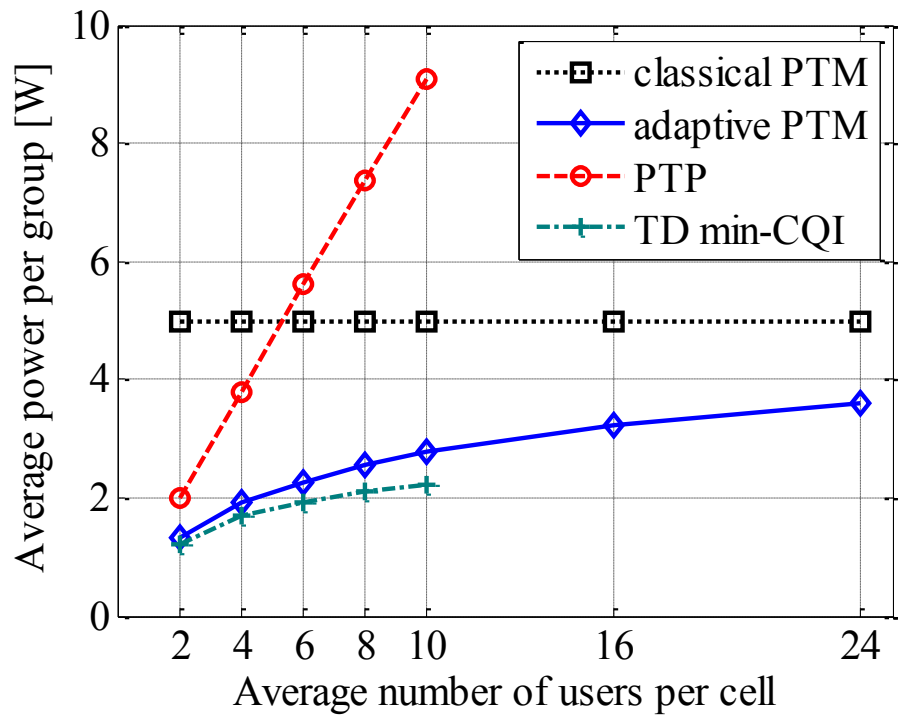

Figure 2. Average power consumption per MBMS user.

With the introduction of feedback channels allowing for adaptive PTM transmission, power gains of 33 % can already be achieved for two users per cell compared to MBMS using PTP connections. The gains increase with the user density. Compared to classical PTM, power gains of 73 % can be achieved with two users per cell. In adaptive PTM mode, the resource consumption per MBMS group gradually increases with the user density because of the growing probability of having users with poor channel quality (see Fig. 1). Thus, the resource savings compared to classical PTM decrease accordingly. The minimum-CQI time-domain scheduling as described in Section A can further enhance the performance of adaptive PTM. However, this scheme is only appropriate for small user groups to limit the impact of increasing delay.

As can be seen in Fig. 3, the USRs for all adaptive PTM schemes are far above the coverage requirements of 95 %. While an average load of 70 % in terms of power is targeted, the load variation between the cells is very large for an average of two MBMS users per cell. Note that this load variation is reflected in a degraded USR. Although increased group sizes allow for improved channel-dependent scheduling in PTP mode, a larger number of MBMS users also results in increased competition, so that packets may be dropped or arrived with larger delay to users with poor radio conditions. With increased group sizes, more unsatisfied users can be observed for the min-CQI time-domain scheduling technique due to a higher probability of having a low aggregate multi-user channel quality.

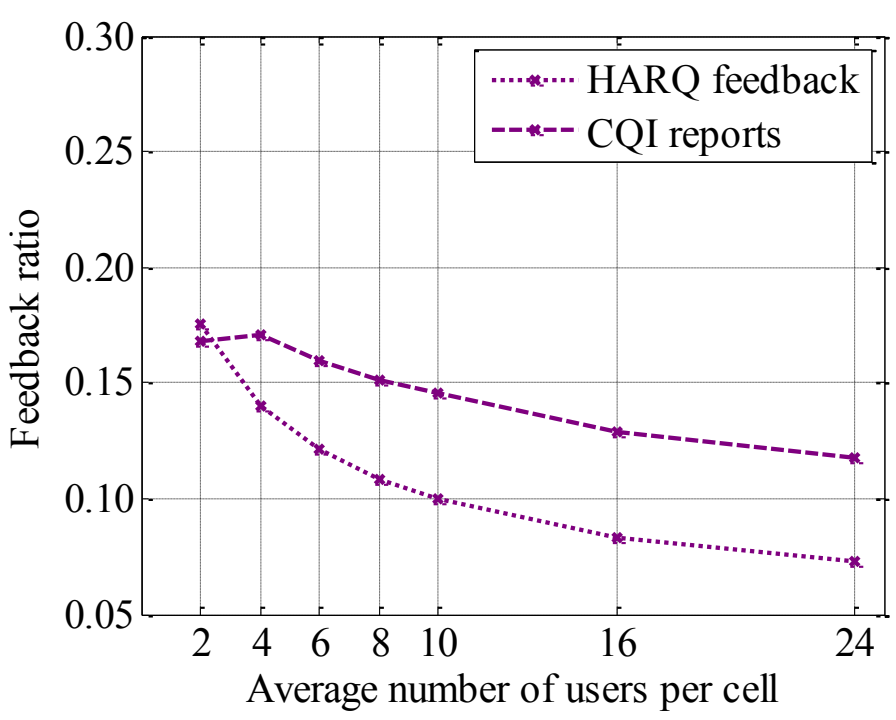

Figure 3. Uplink resource savings by NACK-oriented feedback scheme.

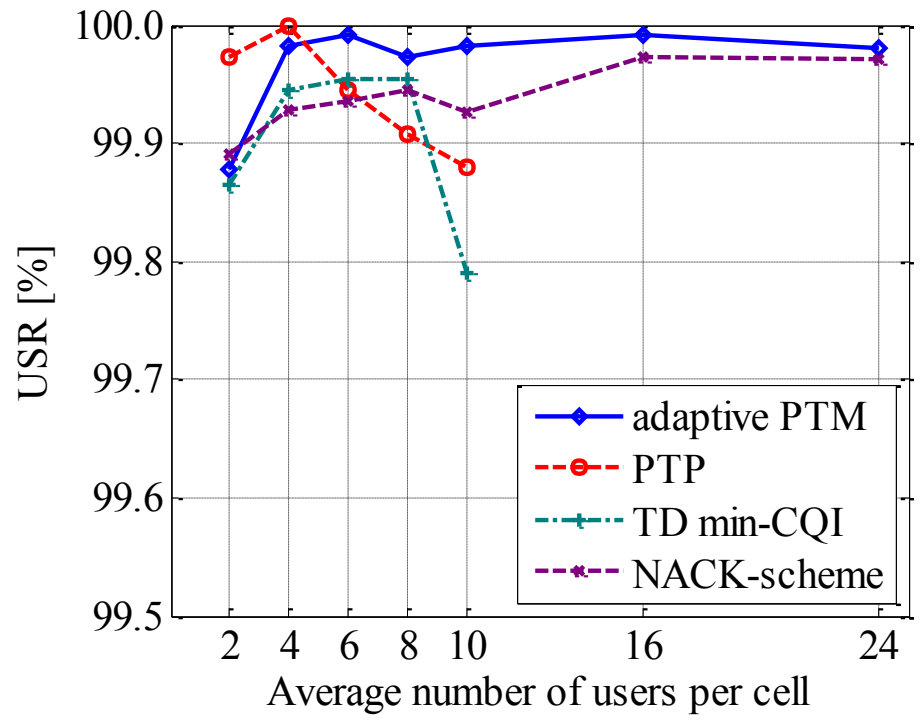

Figure 4. User satisfaction rate.

Using the NACK-oriented feedback scheme described in Section III.C, the amount of uplink load can be significantly reduced compared to the baseline scenario with a combined ACK/NACK scheme and periodic CQI reporting. Fig. 4 shows the feedback ratio of HARQ status and CQI reports. The amount of HARQ status reports is reduced to 18 % for an average of two users per group, and decreases with the group size to 7 % for 24 users per group. The feedback ratio regarding CQI reports show the same trend, but falls with a lower slope. Slow link adaptation to good channel qualities in the NACK-oriented feedback scheme results in higher resource consumption as can be seen in Fig. 5.

Fig. 6 shows the transmit rate per TTI rather than the throughput. For PTP, the transmit rate refers to each single user, while in the PTM transmission modes, it refers to one group. Furthermore, it only takes into account the MCS, but does not distinguish between initial and retransmissions. As classical PTM requires 5 transmissions, the transmit rate for adaptive PTM may be lower, but since less retransmissions are required less resources are used on average.

Fig. 7 shows the average number of HARQ attempts per transport block. 1.1 HARQ attempts on average are needed per user because the target BLER for PTP is 10 % per user. As the same BLER per user is targeted for the baseline configuration of adaptive PTM, the number of necessary HARQ attempts increases with the group size. The NACK-oriented feedback scheme uses conservative MCS selection and can only slowly adapt to better channel conditions. Consequently, the number of HARQ attempts is rather low. Concerning the USR, there are only slight degradations for the NACK-oriented feedback scheme, but anyway above 99.9 % and thus far beyond the requirements of 95 % achieved by classical PTM transmission.

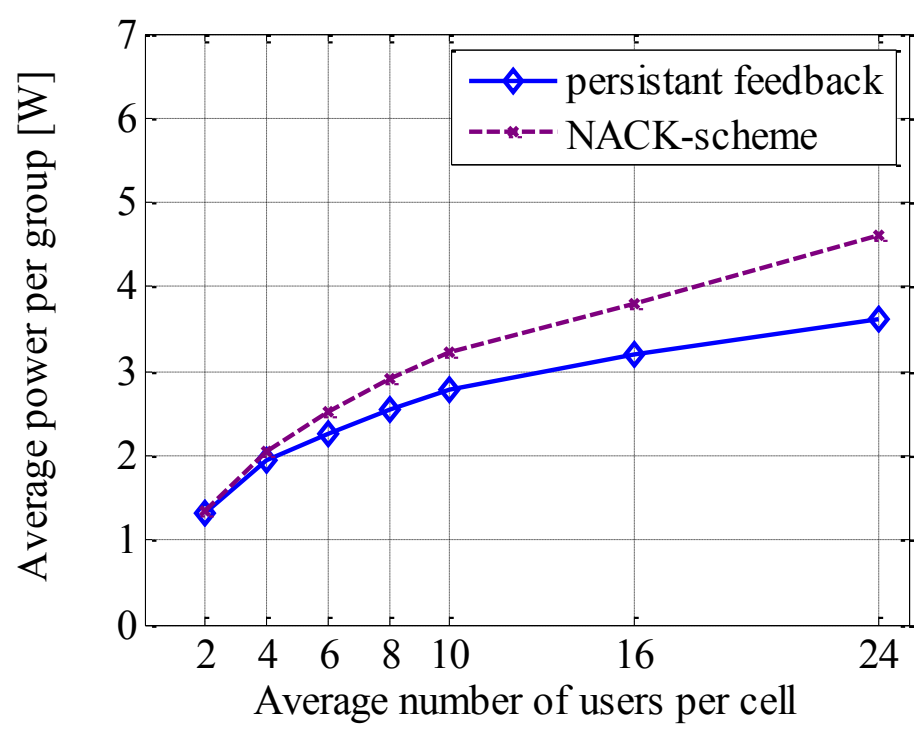

Figure 5. Power consumption of NACK-oriented feedback scheme.

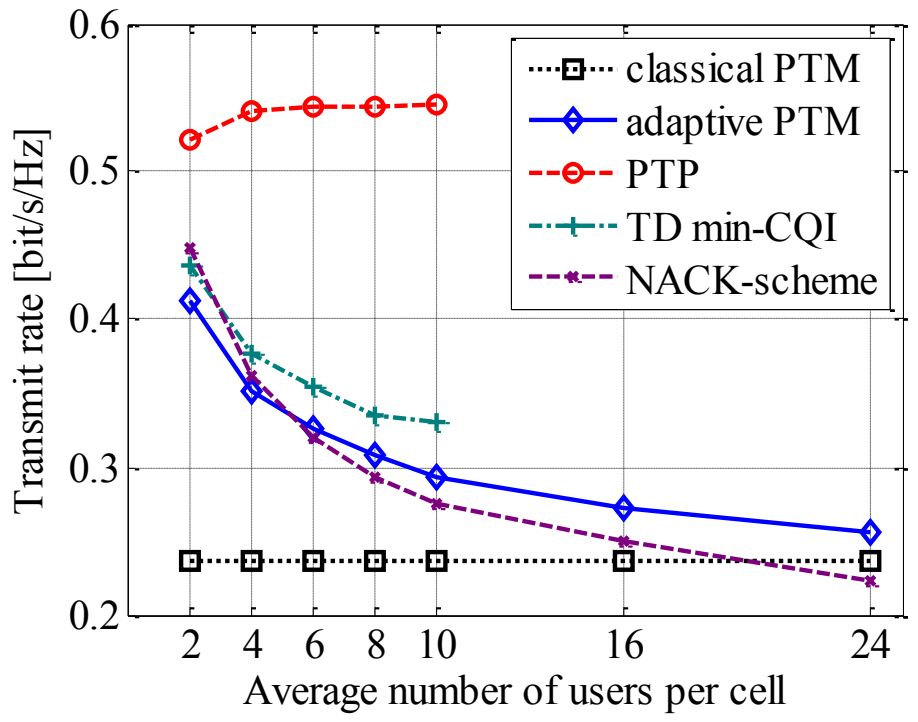

Figure 6. Transmit rate in each scheduled TTI.

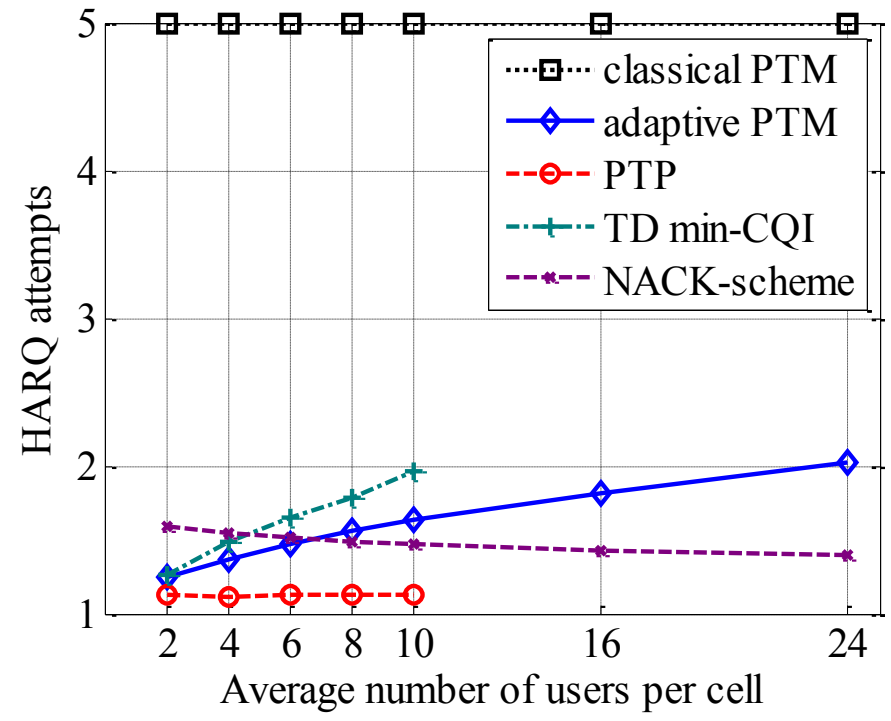

Figure 7. Average number of HARQ attempts per transport block.

## VII. Conclusions

This paper showed that the performance of MBMS in LTE can be enhanced by introducing a PTM transmission mode based on the PTP concept. With an additional feedback channel for each user the radio resource efficiency of PTM transmissions can be improved. Thus, link adaptation for PTM connections can be performed based on both channel quality reports and a packet acknowledgment mechanism. Radio network simulations indicate that at least 33 % of the radio resources can be saved compared to PTP transmission, and opposed to classical PTM transmission without link adaptation, the gains in terms of saved radio resources range from 28 % to 73 % depending on the number of MBMS users interested in the same data service. Further gains can be achieved if the USR is reduced from 99.8 % to 95.0 %. Moreover, a NACK-oriented feedback mechanism to decrease the amount of feedback in the uplink is also presented. The drawback of this technique is that it is not compliant with the PTP transmission mode. However, slightly degraded gains compared to the adaptive PTM transmission with full feedback can be observed while significantly reducing uplink traffic generated by MBMS users.

For several reasons, the application of PTM including feedback is restricted to a small user group. First, the probability that one user within an MBMS group experiences poor channel conditions in a TTI increases with the group size. Second, frequent feedback from receivers transmitting CQI and HARQ status reports on dedicated channels consume a significant amount of the uplink capacity in cells with a large number of users in the MBMS group. The paper confirmed that uplink load can be reduced by using a NACK-oriented feedback scheme.